\documentclass[conference]{IEEEtran}
\usepackage{cite}
\usepackage{amsmath,amssymb,amsfonts}
\usepackage{algorithmic}
\usepackage{graphicx}
\usepackage{textcomp}
\usepackage{xcolor}
\usepackage{enumitem}
\usepackage{tikz}
\usetikzlibrary{shapes, arrows.meta, positioning, calc, decorations.pathreplacing}
\usepackage[margin=1in]{geometry}
\usepackage{booktabs}
\usepackage{hyperref}
\usepackage{listings}

\lstdefinestyle{pythonstyle}{
  basicstyle=\footnotesize\ttfamily,
  keywordstyle=\color{blue!70!black},
  stringstyle=\color{green!40!black},
  commentstyle=\color{gray},
  showstringspaces=false,
  breaklines=true,
  breakatwhitespace=false,
  columns=fullflexible,
  keepspaces=true,
  language=Python,
  frame=single,
  framerule=0.3pt,
  rulecolor=\color{gray!40},
  captionpos=b,              
  abovecaptionskip=4pt,      
  belowcaptionskip=2pt,      
  xrightmargin=0.5em,
  framexrightmargin=0.5em,
}
\def\BibTeX{{\rm B\kern-.05em{\sc i\kern-.025em b}\kern-.08em
    T\kern-.1667em\lower.7ex\hbox{E}\kern-.125emX}}

\begin{document}
\title{ITAS: A Multi-Agent Architecture for LLM-Based Intelligent Tutoring}

\author{\IEEEauthorblockN{Iizalaarab Elhaimeur}
\IEEEauthorblockA{\textit{Center for Real-Time Computing} \\
\textit{Computer Science Department}\\
\textit{Old Dominion University}\\
Norfolk, VA \\
ielha003@odu.edu}
\and
\IEEEauthorblockN{Nikos Chrisochoides}
\IEEEauthorblockA{\textit{Center for Real-Time Computing} \\
\textit{Computer Science and Physics Departments}\\
\textit{Old Dominion University}\\
Norfolk, VA \\
nikos@cs.odu.edu}
}

\maketitle

\begin{abstract}
Large language model tutors are easy to build in a notebook and hard to run in a real course. We describe ITAS (Intelligent Teaching Assistant System), a multi-agent tutoring system that a graduate quantum computing course used for a semester at Old Dominion University. The system has three layers. The teaching layer is a Spoke-and-Wheel of three parallel specialist agents (Video, Code, Guidance) followed by a Synthesizer, plus a separate autograder that evaluates both the correctness and the approach of checkpoint submissions. The operational layer is four Cloud Run microservices with session state in Cloud SQL and interaction events streamed through Pub/Sub to BigQuery. The feedback layer is a narrow-scope conversational agent that answers instructor questions over per-lesson pseudonymized event streams, addressing what we call the Blind Instructor Problem: LLM tutors accumulate more data about students than the instructor can reach through routine channels. The architecture is a direct response to specific failures of an earlier prototype, and we describe which of those fixes carried forward and which were dropped for this iteration. We report on a pilot deployment (five students, one course, one semester) interpreted as system-behavior evidence rather than learning-outcome evidence: the teaching layer handled 334 chat turns without the task-boundary hallucinations that domain consolidation would have risked, the operational layer captured 10,628 events across five modules, and the feedback layer surfaced two findings the instructor acted on mid-semester. We do not claim the pilot generalizes. We do claim that the system as described is one workable answer to the question of what an LLM-based ITS needs to look like end-to-end to run in a real course.
\end{abstract}

\begin{IEEEkeywords}
Intelligent Tutoring Systems, Multi-Agent Systems, Large Language Models, Cloud Microservices, Learning Analytics, Quantum Computing Education, FERPA Compliance
\end{IEEEkeywords}

\section{Introduction}
\label{sec:intro}

Large language models have made intelligent tutoring systems conversational in a way that earlier rule-based tutors could not be. A student can ask a question in natural language and get a contextual response, and deployed systems routinely generate thousands of such interactions per course~\cite{kasneci2023chatgpt,chu2025llmagents}. What the research literature has under-reported is the systems work between ``a single agent that answers questions in a notebook'' and ``a system running in a real course with real students.''

This paper describes one answer to that question. ITAS is a multi-agent tutoring system built for graduate quantum information science education and deployed for a semester at Old Dominion University. This paper focuses on the system itself: the architecture, the infrastructure, the feedback layer, the engineering decisions, and how the design got there from the prototype that preceded it. The domain-specific findings from the same pilot — curriculum, quantum-education outcomes, student engagement patterns — are reported separately~\cite{elhaimeur2026quantum}, as is the latency and cost analysis~\cite{elhaimeur2026latency}.

The system evolved out of an earlier prototype~\cite{elhaimeur2025toward} that failed in specific, diagnosable ways. An initial single Augmented LLM, responsible for planning, teaching, pacing, and tool-calling at once, hallucinated tools and skipped lesson steps. The response was a knowledge-graph-augmented architecture: two specialized LLM agents (a Teaching Agent for real-time interaction and a Lesson Planning Agent for lesson generation) coordinated through a shared knowledge graph that held users, learning resources, interactions, and student state as first-class structured data. A residual class of hallucinations around pacing inference then motivated a student-selectable tag system. None of these changes was sufficient to move the prototype into a real course; it was validated only on simulated runs. The system in this paper is what came next. It carries forward two ideas from the prototype (separation of concerns and multi-agent design) and sets aside two (the knowledge graph and the tag system), for reasons we make precise in Section~\ref{sec:prototype}.

\subsection{Contributions}

This paper makes four contributions.

\begin{enumerate}[leftmargin=*,noitemsep,topsep=2pt]
\item A full description of the ITAS architecture across three layers (teaching, operational, feedback), with design rationale for each component and explicit accounts of what was kept and what was dropped from the prototype.
\item The Spoke-and-Wheel teaching layer: three parallel specialist agents with Pydantic-typed structured outputs, coordinated by a Synthesizer that reads all three reports and produces one natural-language response, plus a separate autograder that enforces both output-correctness and approach-correctness on checkpoints.
\item A feedback layer that addresses the Blind Instructor Problem through a narrow-scope conversational agent over per-lesson pseudonymized event streams. The narrow scope is the reliability argument, applied recursively from the teaching layer.
\item Pilot deployment evidence interpreted as system-behavior validation: the teaching layer did not exhibit the task-boundary failures that consolidating video, code, and conceptual guidance into one agent would have risked, the operational layer captured the interaction volume without loss, and the feedback layer produced findings the instructor acted on. We are explicit that this is a pilot and that the evidence speaks to feasibility of the design, not to learning outcomes.
\end{enumerate}

The paper covers related work, prototype background, the three-layer architecture, pilot evidence, and closes with limitations and future work.
\section{Related Work}
\label{sec:related}

\subsection{Intelligent Tutoring Systems}

Intelligent tutoring research has long established that one-on-one instruction outperforms classroom teaching~\cite{bloom1984twosigma,vanlehn2011effectiveness}, and rule-based cognitive tutors demonstrated that automated instruction can approach one-on-one effectiveness at scale~\cite{anderson1995cognitive,graesser2004autotutor}. Meta-analyses confirm the learning gains but also confirm that sustained multi-session deployments of modern ITS are rare~\cite{kulik2016meta}. The LLM era has renewed interest. Tack and Piech~\cite{tack2022aiteacher} and the BEA 2023 shared task~\cite{tack2023bea} measured the gap between LLM and human pedagogy. Vanzo et al.~\cite{vanzo2024gpt4homework} found learning gains from GPT-4 tutoring in a randomized trial, and LearnLM~\cite{jurenka2024learnlm} showed that pedagogically fine-tuned Gemini is preferred over prompted baselines. Kasneci et al.~\cite{kasneci2023chatgpt} survey opportunities broadly; Chu et al.~\cite{chu2025llmagents} survey LLM agents in education specifically.

The systems dimension is where the gap lies. Yan et al.'s scoping review of 118 papers on LLMs in education~\cite{yan2024practical} explicitly names cost, scalability, and deployment infrastructure as open barriers and does not find a single surveyed work that reports quantitative deployment characteristics under real classroom load. This paper attempts to close part of that gap.

\subsection{Multi-Agent LLM Systems}

Multi-agent frameworks have grown quickly since ReAct~\cite{yao2023react} demonstrated interleaved reasoning and tool use. AutoGen~\cite{wu2024autogen} generalized this to configurable multi-agent conversations. MetaGPT~\cite{hong2024metagpt} encoded standardized operating procedures per role, reducing cascading hallucinations. CAMEL~\cite{li2023camel} explored communicative agents, and CoALA~\cite{sumers2024coala} formalized cognitive architectures for language agents. Guo et al.~\cite{guo2024multiagent} survey the broader space. Educational multi-agent systems include GenMentor~\cite{wang2025genmentor} and SimClass~\cite{zhang2024simclass}; both decompose by task (planning versus execution, or teacher versus student role) rather than by domain.

The most directly relevant recent result is from Kim et al.~\cite{kim2025scaling}, which reports that centralized coordination of specialist agents improves task performance by 80.9\% on parallelizable tasks and contains error amplification to 4.4$\times$ compared with 17.2$\times$ for independent generalist agents. The Spoke-and-Wheel teaching layer in Section~\ref{sec:teaching} is an instance of this centralized-specialist pattern.

\subsection{Learning Analytics for Instructor Feedback}

Behavioral signals have been used as proxies for engagement and confusion for over a decade. Guo et al.~\cite{guo2014video} showed that video seek and drop-off patterns track content difficulty. Blikstein~\cite{blikstein2014programming} demonstrated that code execution telemetry reveals productive struggle invisible in assessment outcomes, and Blikstein and Worsley~\cite{blikstein2016multimodal} extended this to multimodal streams combining video, code, and interaction logs. Kizilcec et al.~\cite{kizilcec2013archetypes} showed that clustering interaction patterns surfaces learner subpopulations that aggregate metrics miss.

The methodology consensus is that cross-modal signals yield insights that single streams miss. The interface consensus, implicitly, is that those insights surface through dashboards. What the literature does not report is an instructor-facing conversational analytics layer over LLM tutoring data that reasons across modalities in response to natural-language instructor questions. The feedback layer in Section~\ref{sec:feedback} is one such design.

\subsection{Privacy in Educational Data}

The Family Educational Rights and Privacy Act~\cite{ferpa1974} constrains any instructor-facing system that has access to student records. LLM tutoring data is harder to anonymize than gradebook data because the content (what a student asked, what error they hit, what they admitted not understanding) can be re-identifying on its own. The design of the feedback layer treats FERPA compliance as an architectural constraint, not a legal overlay: the agent has no runtime path to real identities, because the data it reads is pseudonymized before the agent's callback ever sees it.

\section{Background: The Prototype}
\label{sec:prototype}

This section recaps the prototype's evolution at the depth needed to understand what the current system inherits and what it does not. The prototype is described in full in~\cite{elhaimeur2025toward}. It was never deployed to a real course; it was validated on simulated runs.

\subsection{Five-Step Evolution}

The prototype evolved in five steps, each motivated by a specific failure of the previous step.

\textbf{Step 1: Three-module integrated platform.} Video Player, Code Editor, and Chat Interface were co-located in one UI. No LLM-level intelligence yet, just the substrate.

\textbf{Step 2: Single Augmented LLM.} One LLM was responsible for planning, teaching, pacing, and tool-calling at once. It hallucinated tool calls (calling unnecessary tools, skipping necessary ones), mis-paced lesson advancement, and deviated from the lesson plan. The root cause was cognitive overload: the single agent could not hold all four responsibilities without one interfering with another.

\textbf{Step 3: Two-agent separation of concerns.} A Lesson Planning Agent handled curriculum, and a Teaching Agent handled real-time interaction. Load per agent dropped and hallucinations decreased. A new problem appeared: the two agents had no way to share state or coordinate.

\textbf{Step 4: Knowledge graph as shared state and design substrate.} A graph database held users, learning resources, interactions, and student state as nodes and edges. Both agents read and wrote against the graph. The graph served two roles: it gave the two-agent system a coherent shared memory, and it was designed from the start to support adaptive lesson sequencing by reasoning over interaction history. The KG was the first-class data substrate, not an implementation detail.

\textbf{Step 5: Student-selectable tag system.} The Teaching Agent retained responsibility for inferring when the student was ready to advance. It sometimes got this wrong, missing explicit ``I'm ready'' statements or advancing prematurely. The tag system (Ready, Hint, Media, Confusion) let students declare intent directly rather than requiring the agent to infer it. This cut the residual pacing-inference hallucinations.

\subsection{What This Iteration Kept and What It Dropped}

Two things carried forward from the prototype into the system described in this paper, and two things were set aside.

\textbf{Kept: separation of concerns.} The prototype's key lesson is that one LLM cannot hold too many responsibilities at once. The current system applies this principle more aggressively, decomposing by domain (what the agent reasons about) rather than by task (what it does). Section~\ref{sec:teaching} makes this precise.

\textbf{Kept: multi-agent design.} The current system has four teaching agents instead of two, plus an autograder and a feedback agent. The move is quantitative, not qualitative: more specialization, same underlying principle.

\textbf{Set aside: knowledge graph.} The prototype's KG served two purposes: communication between agents and a substrate for adaptive planning. Neither purpose is load-bearing in the current execution layer. The four Spoke-and-Wheel agents do not require cross-session state to coordinate within a single turn; structured outputs and a short-lived synthesis step are enough. Session-based state in Cloud SQL holds what the KG's state-management role held in the prototype. The adaptive-planning role is genuinely deferred and will return when the planning layer is built.

\textbf{Set aside: tag system.} The tags solved the Teaching Agent's pacing-inference problem. The Spoke-and-Wheel architecture has no Teaching Agent making pacing decisions. The three specialist agents answer questions about the current state; the student advances by submitting checkpoints. The structural problem the tags addressed does not arise in the new design, and we do not plan to reintroduce them.

\section{System Overview}
\label{sec:overview}

\begin{figure}[t]
\centering
\begin{tikzpicture}[
    scale=0.72, transform shape,
    every node/.style={font=\footnotesize},
    layer/.style={rectangle, draw, rounded corners=4pt, minimum height=1.6cm,
                  minimum width=7.0cm, align=center, line width=0.6pt,
                  fill=#1!12, draw=#1!60},
    svc/.style={rectangle, draw, rounded corners=3pt, minimum height=0.7cm,
                minimum width=1.7cm, align=center, font=\tiny,
                fill=#1!15, draw=#1!60, line width=0.4pt},
]
\node[layer=blue, label={[font=\small]left:Teaching Layer}] (teach) at (0,3.2)
    {Spoke-and-Wheel: Video + Guidance + Code $\rightarrow$ Synthesizer\\
     Autograder (separate agent, checkpoint evaluation)};

\node[layer=orange, label={[font=\small]left:Operational Layer}] (ops) at (0,0)
    {Four Cloud Run microservices\\
     Session state in Cloud SQL, events in BigQuery via Pub/Sub};

\node[layer=green!60!black, label={[font=\small]left:Feedback Layer}] (fb) at (0,-3.2)
    {Conversational analytics agent\\
     Pseudonymized per-lesson event streams + lesson metadata};

\draw[-{Stealth[length=4pt]}, thick, gray] (teach.south) -- (ops.north)
    node[midway, right, font=\tiny, color=gray] {events};
\draw[-{Stealth[length=4pt]}, thick, gray] (ops.south) -- (fb.north)
    node[midway, right, font=\tiny, color=gray] {pseudonymized};
\end{tikzpicture}
\caption{The three layers of ITAS. The teaching layer runs in real time during student interaction. The operational layer captures events and persists state. The feedback layer lets the instructor query aggregated patterns conversationally. Arrows denote the data flow that binds the three layers into one system.}
\label{fig:overview}
\end{figure}
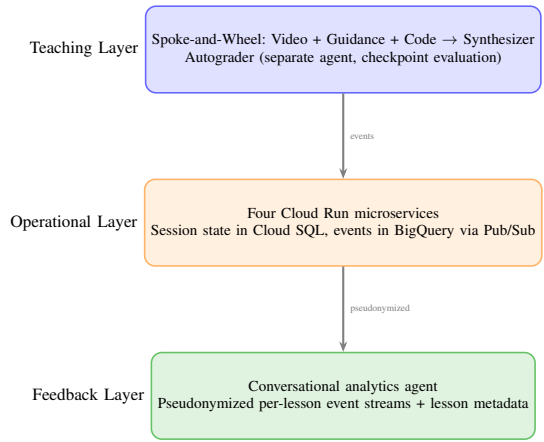

ITAS has three layers (Fig.~\ref{fig:overview}).

The \textbf{teaching layer} runs during student interaction. It receives a student's chat question plus their current lesson state (notebook cells, code, outputs, video transcript, checkpoint) and returns a single natural-language response. It consists of the four-agent Spoke-and-Wheel (three parallel specialists plus a Synthesizer) and a separate autograder invoked when the student submits a checkpoint.

The \textbf{operational layer} is the infrastructure. It exists so the teaching layer can run at classroom scale and so every interaction gets recorded for later. Four Cloud Run microservices handle teaching, code execution, analytics ingestion, and autograding. Session state lives in Cloud SQL (Postgres) so that progress survives across sessions and agents see consistent context. Every student action generates an event that streams through Pub/Sub to BigQuery under a pseudonymized identifier.

The \textbf{feedback layer} is what the instructor sees. A single conversational agent reads per-lesson event streams (pseudonymized) plus lesson metadata, and answers instructor questions in natural language over that context. It does not write SQL at runtime. It does not compute new statistics. It does not have access to real student identities. Its only job is to narrate what the callback puts in front of it.

The next three sections describe each layer in detail.

\section{Teaching Layer: Spoke-and-Wheel}
\label{sec:teaching}

\subsection{Design Rationale}

The prototype ended up with two agents plus a knowledge graph plus a tag system, and the Teaching Agent was still broadly responsible for real-time student interaction during lesson execution. In preparing to scale from simulated runs to a real course covering the full QIS domain, we expected that consolidating video comprehension, code debugging, and conceptual guidance into a single agent prompt would reintroduce the kind of cognitive overload the prototype's earlier single-agent form had exhibited (hallucinated tool calls, pacing errors, and abstraction-level mixing). The Spoke-and-Wheel design is a preemptive architectural choice against those failure modes rather than a fix for hallucinations the two-agent prototype itself documented.

The fix was specialization by domain, not by task. Task decomposition (``this agent plans, that agent teaches'') was what the prototype did; it splits \emph{what the agent does} but does not split \emph{what the agent reasons about}. Domain decomposition splits the reasoning surface itself. Each agent has a narrower scope, a narrower context window, and a narrower failure mode. An agent that only reasons about video cannot hallucinate code; an agent that only reasons about code cannot invent timestamps.

\subsection{The Three Specialist Agents}

Three domain boundaries emerged as cognitively distinct: media comprehension, code implementation, and conceptual understanding.

The \textbf{Video Agent} receives the student query, the active checkpoint title, and a timestamp-indexed transcript of the current lecture segment. It identifies which segments address the question, extracts the key insight shown, and notes gaps where the video does not cover what the student needs. It does not reason about code or conceptual correctness; those are not in its scope.

The \textbf{Code Agent} receives the student query, the active checkpoint, the editor language, a Qiskit error catalog in its instruction template, and the student's current code from all cells along with their outputs. It operates in two modes. If code exists and fails, it diagnoses: identifies the error, its location (cell ID and approximate line), and its cause. If code is empty or incomplete, it provides implementation guidance: what the next step is. In both modes it is explicitly forbidden from writing complete solutions.

The \textbf{Guidance Agent} receives the student query, the active checkpoint, the editor language, and a per-lesson instruction template. It identifies what the student fundamentally misunderstands, recommends a pedagogical approach (Socratic question, analogy, decomposition), and flags common misconceptions. It does not write code.

Each agent produces a structured output validated by a Pydantic schema (Listing~\ref{lst:schemas}). The schemas are short on purpose. Early versions had ten-plus fields per agent; we trimmed them to three or four because longer schemas encouraged the agents to produce filler rather than to say only what they knew.

\begin{figure}[t]
\begin{lstlisting}[style=pythonstyle, caption={Pydantic output schemas for the three specialist agents.}, label={lst:schemas}]
class VideoReport(BaseModel):
    relevant_segment: str
    key_insight: str
    coverage_gap: str

class GuidanceReport(BaseModel):
    conceptual_gap: str
    pedagogical_approach: str
    misconception_flag: str

class CodeReport(BaseModel):
    diagnosis: str
    correct_components: str
    next_step: str
    alternative_approach: str
\end{lstlisting}
\end{figure}

\subsection{Why Three Agents, Not Two or Four}

The three-agent count is empirical. Two agents (combining video with one of code or guidance) conflated reasoning domains and re-introduced the cross-domain hallucinations the split was meant to prevent. Four agents (splitting code into syntax-debugging and algorithmic-guidance, for example) added coordination overhead in the Synthesizer without reducing observed failures. Three is what the design converged on, not what theory prescribed.

\subsection{The Synthesizer}

The three specialists run in parallel. A Synthesizer Agent then reads all three structured reports and produces one natural-language response for the student. Its prompt has four elements worth naming.

First, an \textbf{environment constraint} block. Students work in a web notebook, not a local IDE. They cannot open a terminal, install packages with pip, or use a debugger. The Synthesizer is explicitly told not to suggest any of these. Without this, Gemini defaults produce plausible but impossible advice (``open a terminal and run \ldots'').

Second, a \textbf{priority hierarchy}. Specific code errors are addressed first, conceptual gaps second, video references third. The ordering reflects student need: a broken cell is more urgent than a conceptual refinement.

Third, a \textbf{response-format constraint}. The student-facing output is one to four sentences in plain prose, no bullets or headers, with cell IDs referenced inline when code is discussed, and a single concrete next action at the end (``run the cell,'' ``change X to Y'').

Fourth, a \textbf{solution-withholding constraint}. The Synthesizer, like the Code Agent, is forbidden from writing complete solutions. It confirms what the student got right, pinpoints what is wrong, and suggests the next step.

The end-to-end latency of this pipeline is dominated by $\max(L_\text{video}, L_\text{guidance}, L_\text{code}) + L_\text{synth}$. The companion paper~\cite{elhaimeur2026latency} analyzes this in detail, including the parallel-phase maximum effect and throughput-tier implications. All four agents use Gemini 2.5 Flash on Vertex AI with thinking disabled (\texttt{thinking\_budget=0}) to keep latency low.

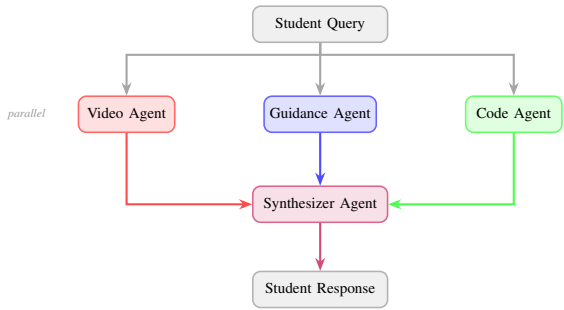
\begin{figure}[t]
\centering
\begin{tikzpicture}[
    scale=0.64, transform shape,
    node distance=1.1cm and 1.4cm,
    every node/.style={font=\small},
    block/.style={rectangle, draw, rounded corners=3pt,
                  minimum height=0.75cm, minimum width=2.0cm,
                  align=center, fill=#1!12, draw=#1!60, line width=0.6pt},
    block/.default=blue,
    arrow/.style={-{Stealth[length=5pt]}, thick, color=#1!70},
    arrow/.default=black,
]
\node[block=gray, minimum width=2.8cm] (query) {Student Query};
\node[block=red,  below left=1.1cm and 1.6cm of query] (video)    {Video Agent};
\node[block=blue, below=1.1cm of query]                 (guidance) {Guidance Agent};
\node[block=green,below right=1.1cm and 1.6cm of query] (code)     {Code Agent};
\node[block=purple,below=1.1cm of guidance, minimum width=2.8cm] (synth) {Synthesizer Agent};
\node[block=gray, below=1.0cm of synth,    minimum width=2.8cm] (response) {Student Response};
\draw[arrow=gray] (query.south) -- ++(0,-0.25) -| (video.north);
\draw[arrow=gray] (query.south) -- (guidance.north);
\draw[arrow=gray] (query.south) -- ++(0,-0.25) -| (code.north);
\draw[arrow=red]  (video.south)    |- (synth.west);
\draw[arrow=blue] (guidance.south) -- (synth.north);
\draw[arrow=green](code.south)     |- (synth.east);
\draw[arrow=purple](synth.south)   -- (response.north);
\node[font=\scriptsize\itshape, color=gray!80, left=0.55cm of video] {parallel};
\end{tikzpicture}
\caption{Spoke-and-Wheel teaching architecture. Three specialist agents run in parallel, each scoped to one reasoning domain. The Synthesizer reads all three structured reports and produces a single natural-language response. End-to-end latency is dominated by the maximum of the three parallel agents plus the synthesizer.}
\label{fig:spokewheel}
\end{figure}

\subsection{The Autograder}

Checkpoint submissions go through a separate agent with its own structured output (Listing~\ref{lst:autograder}). This is the Autograder. It is separate from the Spoke-and-Wheel pipeline because its job is different: not to help the student, but to decide whether a submission passes.

\begin{figure}[t]
\begin{lstlisting}[style=pythonstyle, caption={Autograder output schema. A boolean plus the reasoning that led to it.}, label={lst:autograder}]
class BooleanResponse(BaseModel):
    passed: bool
    reasoning: str
\end{lstlisting}
\end{figure}

The Autograder receives per-checkpoint grading instructions injected at call time, the editor language, and the student's code and outputs for the target cells. Its grading criteria explicitly enforce the dual output-and-approach requirement documented in the companion paper~\cite{elhaimeur2026quantum}: correct output alone is not sufficient; the implementation must also satisfy the approach criteria. In quantum computing this distinction matters because multiple different implementations can produce identical measurement statistics while reflecting different levels of understanding. The Autograder accepts any implementation that satisfies both criteria and rejects correct-output submissions that bypass the underlying concept.

A before-agent callback runs on each submission to short-circuit empty-submission cases without invoking the model. This is a small but non-trivial optimization: the empty case happens routinely (a student clicks Check Progress before writing any code), and handling it in the callback avoids spending a model call on it.

\subsection{Implementation}

The teaching layer is implemented on Google's Agent Development Kit (ADK)~\cite{googleadk}. A \texttt{ParallelAgent} wraps the three specialists, and a \texttt{SequentialAgent} composes the parallel group with the Synthesizer. All four agents are Gemini 2.5 Flash on Vertex AI. Temperature settings are per-agent: the Code Agent uses 0.2 (we want determinism for error diagnosis), the Guidance Agent 0.4 (slight variation for pedagogical phrasing), the Synthesizer 0.5 (natural conversational tone), and the Video Agent 0.3. These values are defaults we did not systematically tune; they worked well enough in the pilot that we left them alone.

Lesson-specific content (video transcripts, checkpoint instructions, agent guidance instructions, debugging instructions) is injected into prompts via placeholders at call time. The agent definitions themselves are lesson-agnostic, so the same teaching layer supports any lesson whose content has been authored in the expected schema.

\section{Operational Layer}
\label{sec:operational}

\subsection{From Local Prototype to Cloud Deployment}

The prototype ran on one machine for one user at a time. That was adequate for simulated validation and completely inadequate for a real classroom. A cloud deployment had to handle concurrent users with isolated sessions, survive failures without dropping student work, sandbox code execution so student code could not affect the host or other students, and store data in a way that met FERPA constraints.

We chose Google Cloud Platform and Cloud Run for the compute. The reason was iteration speed: Cloud Run auto-scales without managing instances, deploys from a container in minutes, and returns to zero when idle so there is no capacity-planning overhead in early development. We did not evaluate GKE, App Engine, or Vertex Agent Engine against these criteria. The decision was ``the simplest thing that would let us deploy and iterate fast.''

\subsection{Four Microservices}

ITAS runs as four Cloud Run microservices, each with a single responsibility.

The \textbf{Teaching Agent Service} hosts the Spoke-and-Wheel pipeline. It is a FastAPI app wrapping the ADK agent definitions, exposing a \texttt{/run} endpoint that accepts a session identifier and a user message. Auto-scaling is configured with a minimum instance count to avoid cold-start latency on the first request of a session.

The \textbf{Python Execution Service} runs student code in a sandboxed Python environment pre-loaded with quantum computing dependencies (NumPy, Qiskit, Matplotlib, SciPy). It exposes \texttt{/execute} (run code in a session namespace), \texttt{/health}, \texttt{/info}, and \texttt{/session/\{id\}/reset}. Variables persist within a session namespace so students can build up state across cells, and namespaces are isolated between students. Code timeouts and memory limits apply per execution; the quantum kernel needs a heavier allocation than a general-purpose sandbox would because circuit simulation is both compute- and memory-intensive.

The \textbf{Analytics Ingestion Service} accepts JSON events from the frontend (video playback, code execution, checkpoint submissions, chat messages, session lifecycle) and publishes them to a Pub/Sub topic that feeds BigQuery. The frontend posts events fire-and-forget: analytics failures are logged but never propagate to the student-facing UI.

The \textbf{Autograding Service} wraps the autograder agent (Section~\ref{sec:teaching}). It is structurally similar to the Teaching Agent Service, with its own \texttt{/run} endpoint and its own per-session state.

\subsection{Session State}

Session state is the one piece of infrastructure that did not fit the stateless-microservice pattern. Three things forced it out: the teaching agents need consistent notebook context across turns, the student needs their work to persist across sessions, and the autograder needs to see the same code the teaching agents saw. We externalized session state to Cloud SQL (Postgres) through ADK's built-in session service, keyed by composite \texttt{session\_\{user\_id\}\_\{lesson\_id\}} identifiers.

Session state held in this path includes notebook cell contents for editable cells, completed-checkpoint flags, and the chat-state context assembled for each teaching-agent turn. It does not include the raw BigQuery event stream (that flows separately to the analytics store).

This is the place to address the knowledge graph honestly. The prototype's KG was designed for two purposes: to coordinate state between the Lesson Planning Agent and the Teaching Agent, and to support adaptive lesson sequencing by reasoning over interaction history. The current execution layer needs neither. The four Spoke-and-Wheel agents coordinate through Pydantic-typed structured outputs within a single synthesis step, not across turns, so they do not need a persistent graph. Session state in Postgres is sufficient for cross-turn continuity. The adaptive-sequencing use case is genuinely deferred; it will return when we build the planning layer.

\subsection{Data Flow and Pseudonymization}

The frontend generates an event for every meaningful student action: cell edits, cell executions, checkpoint attempts, chat messages, video play/pause/seek, session start and end. Events are posted to the Analytics Ingestion Service, which publishes them to Pub/Sub. A streaming sink loads them into BigQuery in near-real-time.

Pseudonymization happens at ingestion. Student identifiers in BigQuery are pseudonyms that do not resolve to real identities without access to the authentication store, which sits outside the BigQuery project. The analytics agent's BigQuery queries (Section~\ref{sec:feedback}) read these pseudonyms; the agent never sees a real name or email. The decoupling is deliberate: an authentication-system compromise does not compromise the analytics store, and a compromise of either does not compromise both.

\subsection{FERPA Compliance as Architectural Constraint}

Compliance with FERPA~\cite{ferpa1974} is enforced in three places that the rest of the system inherits rather than re-implements. First, interaction data is stored under pseudonymized identifiers at ingestion. Second, the feedback agent reads only the pseudonymized event streams and lesson metadata, never the identity-mapping store. Third, all data remains within the institution's controlled GCP environment under existing institutional data governance. The agent does not filter personally identifiable information at query time because the data it can reach never contained any.

\subsection{Cost and Latency}

Under Priority PayGo on Vertex AI, end-to-end response latency stays flat around 3.5--4.0 seconds from one to fifty concurrent users, and per-student cost per semester is below a typical STEM textbook under a worst-case usage ceiling. The companion paper~\cite{elhaimeur2026latency} reports these results in detail, including the parallel-phase maximum effect and throughput-tier tradeoffs. We cite these figures here to close the question of whether the design is economically viable in a real classroom: under the configurations evaluated, it is.

\section{Feedback Layer: Conversational Analytics}
\label{sec:feedback}

\subsection{The Blind Instructor Problem}

An LLM tutoring system accumulates detailed knowledge of each student's trajectory. The instructor running the course sees almost none of it through the channels they are used to reading. This is not a tooling oversight. It is a design tension.

Students may ask an AI tutor questions they would not ask a professor, and admit confusions they would not admit in lecture. Patton's treatment of qualitative inquiry documents a related observation effect: a channel that people know is observable tends to produce different data than one they believe is private~\cite{patton2002qualitative}. To the extent that effect applies here, routinely exposing raw transcripts under real identities risks changing what students are willing to ask in the first place, which would destroy part of what makes the data valuable.

We call the resulting gap the \textbf{Blind Instructor Problem}. Static dashboards help but we think they do not resolve it. A dashboard displays metrics its designer anticipated, and the designer brings priors about what students should struggle with. Learning analytics research has long acknowledged that the choice of what to measure shapes what can be seen, which is why multimodal frameworks~\cite{blikstein2016multimodal} and cross-stream clustering~\cite{kizilcec2013archetypes} exist in the first place. Dashboards also make cross-referencing manual: drop-off at minute 42, a checkpoint failure spike, and a cluster of chat questions about tensor products live in three panels, and the join happens in the instructor's head. And if a dashboard surfaces something interesting, the instructor's follow-up question is usually ``why,'' which a dashboard cannot answer.

The feedback layer in this paper is one design for the missing component. We do not claim it is the only one.

\subsection{Design Constraints}

We take a feedback layer to be adequate for this problem if it satisfies three constraints at once.

\textbf{Pseudonymous.} The agent's context contains no real identities. Students are identified by pseudonyms that are not linkable to names without separate access.

\textbf{Cross-modal.} Video, chat, code, and checkpoint data are reasoned about together within one context, not shown in separate panels.

\textbf{Conversational.} The instructor asks in natural language and follows up without learning a schema or a query language.

\subsection{The Agent}

The feedback layer is a single conversational LLM agent. Its only job is to answer instructor questions in natural language over a context that is assembled for it at the start of each turn. It does not write SQL at runtime. It does not compute new statistics. It does not fetch additional data beyond what the context-assembly step puts in its prompt. It receives the assembled context (per-lesson event streams under pseudonymized identifiers, plus lesson metadata, plus a small summary block) and the instructor's question, and it produces a natural-language answer.

The narrowness is the reliability argument. An agent with no query-writing tool cannot write a bad query. An agent with no computation step cannot compute wrong. An agent with no access to the identity mapping cannot surface real names. These are architectural properties, not prompt-level constraints, which matters because prompt-level constraints on LLMs are known to be bypassable~\cite{huang2024hallucination}. This is the same design principle as the teaching layer, applied recursively at the analytics layer.

\subsection{Pipeline}

\begin{figure}[t]
\centering
\begin{tikzpicture}[
    scale=0.62, transform shape,
    node distance=1.0cm and 1.2cm,
    every node/.style={font=\small},
    block/.style={rectangle, draw, rounded corners=3pt,
                  minimum height=0.75cm, minimum width=2.4cm,
                  align=center, fill=#1!12, draw=#1!60, line width=0.6pt},
    arrow/.style={-{Stealth[length=5pt]}, thick, color=#1!70},
    arrow/.default=black,
]
\node[block=gray, minimum width=3.0cm] (query) {Instructor Query};
\node[block=orange, below=1.0cm of query, minimum width=3.0cm] (agent)
      {Analytics Agent};
\node[block=blue, below left=1.0cm and 1.4cm of agent]  (bq)
      {BigQuery\\(per-lesson events,\\pseudonymized)};
\node[block=green, below right=1.0cm and 1.4cm of agent] (meta)
      {Lesson Metadata};
\node[block=gray, below=2.2cm of agent, minimum width=3.0cm] (response)
      {Natural Language Response};
\draw[arrow=gray] (query.south) -- (agent.north);
\draw[arrow=orange] (agent.south) -- ++(0,-0.3) -| (bq.north);
\draw[arrow=orange] (agent.south) -- ++(0,-0.3) -| (meta.north);
\draw[arrow=blue] (bq.south) |- (response.west);
\draw[arrow=green] (meta.south) |- (response.east);
\end{tikzpicture}
\caption{The feedback-layer pipeline. A pre-agent callback queries BigQuery for the selected lesson's events and joins them with lesson metadata. The agent reads the assembled context plus the instructor's question and produces a natural-language answer. The agent does not issue queries at runtime.}
\label{fig:feedback}
\end{figure}
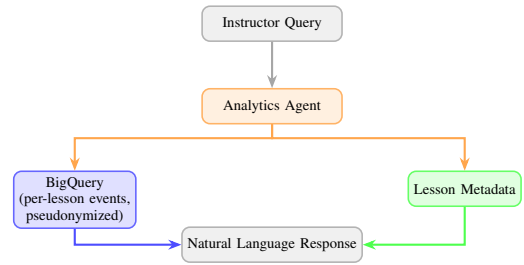

Figure~\ref{fig:feedback} shows the three-stage pipeline.

\textbf{Stage 1: Data pull.} When the instructor selects a lesson, a pre-agent callback runs a fixed set of BigQuery queries scoped to that lesson. The queries are parameterized by lesson identifier but their structure is hard-coded in the callback. The agent runtime has no SQL or BigQuery tool registered; it cannot issue queries at runtime regardless of prompt content. The query set returns per-lesson event streams: chat messages (sender, timestamp, message text truncated at 300 characters), video playback events (play, pause, seek, with before-and-after timestamps on seeks), and code and checkpoint events (cell identifier, pass/fail status, error message for failures, checkpoint reasoning for checkpoint outcomes). A final query returns a small summary block (total students, total sessions, total counts per event type for the lesson). All streams come back under pseudonymized identifiers.

\textbf{Stage 2: Context assembly.} The callback formats the query results into a plain-text document grouped by pseudonymized student and joins it with lesson metadata: learning objectives, notebook structure (which cells are editable, which are checkpoints), and the video outline. The metadata is what lets raw event counts become pedagogically interpretable. A cluster of seeks around minute 42 is a number; paired with metadata saying ``minute 42 is the transition from single-qubit to multi-qubit states,'' it becomes readable as a specific concept students are revisiting.

\textbf{Stage 3: Narration.} The assembled context and the instructor's question go to the LLM. It reads the context and answers in natural language. If the context does not contain what the instructor asked about, the agent is instructed to say so rather than speculate. For multi-turn sessions, prior exchanges are cached and included in subsequent prompts, so the instructor can build on earlier questions without re-establishing context.

\subsection{Prompt Design}

Two prompt-level choices matter enough to name.

First, the agent is instructed to write in natural prose, not bullets or markdown headers. The instruction template is explicit: ``talk like a person, not a report generator.'' This is a reaction to the default tendency of Gemini 2.5 Flash to produce heavily-formatted responses with headers and numbered recommendations. For instructor analytics, heavy formatting reads as canned output; prose reads as someone who actually looked at the data.

Second, the agent is instructed to cross-reference across data types when it tells a story. The instruction template gives an example of the pattern we want: a student who barely watched the video and then immediately struggled with the coding exercise is a single story, not two observations. This kind of cross-modal reasoning is the design's reason for existing, and it has to be prompted explicitly because the model's default is to narrate each stream in turn.

\section{Pilot Deployment}
\label{sec:deployment}

ITAS ran in a graduate quantum computing course at Old Dominion University for one semester in a flipped-classroom model. The pilot was five students, one instructor. All data is stored under pseudonymized identifiers per the operational layer. Domain-specific findings about QIS education are reported separately~\cite{elhaimeur2026quantum}; we use a subset of the same numbers here and interpret them as system-behavior evidence.

\subsection{Evidence Base}

Table~\ref{tab:events} summarizes the interaction volume. These figures come from the same BigQuery store the feedback agent reads; the companion paper~\cite{elhaimeur2026quantum} reports them in a different interpretive frame.

\begin{table}[h]
\centering
\caption{Pilot Deployment Interaction Volume}
\label{tab:events}
\begin{tabular}{lr}
\toprule
\textbf{Event Category} & \textbf{Count} \\
\midrule
Video playback events       & 7{,}666 \\
Chat messages (student + AI) & 334 \\
Code executions              & 387 \\
Code execution success rate  & 77\% \\
Code editor interactions     & 147 \\
Session management events    & 124 \\
Checkpoint evaluations       & 32 \\
Error events                 & 208 \\
\midrule
\textbf{Total events}        & \textbf{10{,}628} \\
\bottomrule
\end{tabular}
\end{table}

\subsection{Teaching Layer}

Across 334 chat turns, the teaching layer handled the full range of student questions (mathematical formalism, code debugging, video-referenced concept questions, cross-module synthesis) without the kind of task-boundary hallucinations that consolidating those reasoning domains into a single agent would have risked: the Video Agent did not suggest timestamps when reasoning about code-only problems, the Code Agent did not misinterpret timestamps as function references, and the Synthesizer did not conflate abstraction levels when moving between formalism and implementation. We are careful about the claim. We did not run a head-to-head comparison against a single-agent baseline or against the prototype. What we can honestly say is that the failures domain consolidation would have risked are not the failures the current system made. This is consistent with Kim et al.'s scaling principle that specialization contains error amplification~\cite{kim2025scaling}, but consistent-with is not proof.

A second observation is about mathematical depth. Graduate QIS students pushed the Guidance Agent into group-theoretic definitions and cross-module synthesis over the course of the pilot. The companion paper~\cite{elhaimeur2026quantum} reports on this as a finding about graduate-level engagement with AI tutors. From a system perspective, it suggests that the agent's instruction calibration (``These are graduate students. Engage with the mathematics; do not oversimplify.'') was doing real work, and that the instruction template is not cosmetic.

\subsection{Operational Layer}

The infrastructure ran for the full semester without a system-wide failure. Cloud Run auto-scaled as expected. Session state persisted across students and modules: a student returning the following week saw their previous work intact. The Pub/Sub-to-BigQuery analytics pipeline captured every event; the fire-and-forget design meant no analytics incident affected the student-facing path. The companion paper~\cite{elhaimeur2026latency} reports latency and cost behavior over the same deployment window.

\subsection{Feedback Layer}

The feedback agent surfaced two findings the instructor acted on mid-semester. The companion paper~\cite{elhaimeur2026quantum} describes both in the context of QIS pedagogy; here we describe them as evidence of the feedback layer working as designed.

The \textbf{dead zone finding}: asked about Module~2 engagement, the agent reported a cluster of students who stopped watching the lecture around the 42-minute mark. Cross-referencing the exercise-coverage metadata, it reported that all four Module~2 checkpoints cover single-qubit operations while the lecture content from minute~44 onward covers multi-qubit states. It framed this as a coverage gap: students watched up to the point where the assessment incentive ran out. The instructor revised the Module~2 exercise set to include multi-qubit checkpoints.

The \textbf{confusion-versus-carelessness distinction}: asked whether a specific checkpoint failure reflected conceptual misunderstanding, the agent distinguished a submission using an incorrect variable name (reading comprehension issue: the instructions specified a name, the student used a different one) from a submission hitting a \texttt{QiskitError} because of using Python's \texttt{*} operator where \texttt{numpy.dot} was needed (implementation-level knowledge gap). The two submissions had identical pass/fail status in the gradebook; the pedagogical response they called for was different.

What the two findings have in common is that they required cross-referencing across data types and they required interpretive follow-up. Neither came out of a single dashboard panel. Both were surfaced during conversations the instructor initiated with natural-language questions, not by the agent writing new queries or computing new statistics. No real student identity was exposed to the instructor through the feedback layer.

\section{Limitations}
\label{sec:limitations}

The pilot is small: five students, one course, one semester, one instructor. We do not have a control condition, we do not have a dashboard-only comparison, and we do not have learning-outcome data. The evidence in Section~\ref{sec:deployment} is feasibility evidence, not effectiveness evidence.

Several more specific limitations are worth naming.

\textbf{No controlled comparison against a single-agent baseline or against the prototype.} We claim that the teaching layer avoided task-boundary failures that domain consolidation would have risked. We did not construct a single-agent baseline or replay prototype conditions against the current system. The claim is diagnostic, not comparative.

\textbf{The instructor designed both the curriculum and the system.} Self-selection bias cuts both ways: the curriculum was designed for the system and the system was designed for the curriculum. The companion paper~\cite{elhaimeur2026quantum} names this explicitly as a limitation of the deployment, and it applies to this paper as well.

\textbf{The feedback agent's narration is not formally evaluated.} We did not compare its interpretations against expert raters who had seen the same data. Informal review by the instructor suggested the interpretations were reasonable and actionable. A rigorous evaluation against expert baselines is future work.

\textbf{Pseudonymous per-student streams are a weaker privacy property than full aggregation.} The agent's context contains per-student event sequences under pseudonyms, not aggregates across students. An instructor reading the agent's output could track one pseudonym's behavior across modalities within one session. We think this is acceptable for a single-course pilot where the pseudonyms do not link to real identities without separate access, but the tradeoff is worth naming. A production deployment at larger scale should revisit whether full aggregation is preferable.

\textbf{The system was not red-teamed against prompt injection or adversarial queries.} The architectural guarantees (no SQL writing, no identity-mapping access) hold by construction. We did not attempt to circumvent them through adversarial instructor queries or prompt injection via student-authored content.

\textbf{Generalization beyond STEM flipped classrooms is untested.} The metadata structure that makes the feedback layer useful (checkpoint-to-video alignment, error-to-concept mapping) is particular to a flipped-classroom course with explicit assessment checkpoints. Subjects with different assessment structures would need different metadata, and we have not worked out what that looks like.

\section{Future Work}
\label{sec:future}

\textbf{The planning layer and the knowledge graph.} This paper describes the execution layer only. A planning layer would handle curriculum authoring, adaptive sequencing based on student behavior, and cross-session pedagogical strategy. This is where the prototype's knowledge graph~\cite{elhaimeur2025toward} returns as a candidate design: the graph's adaptive-sequencing use case is unchanged, and the state-reasoning infrastructure it provides is a natural fit for planning-layer work. The graph's execution-layer roles (real-time state management, cross-turn coordination) are covered by session state in Postgres and do not need the graph.

\textbf{Proactive feedback.} The feedback layer is reactive: the instructor has to ask. A proactive version would run the cross-references automatically and surface patterns without waiting. The dead zone finding would have been more useful earlier in the semester if it had been surfaced proactively. Alert fatigue is the obvious risk, and any proactive version would need careful thresholding.

\textbf{Formal evaluation of both the teaching and feedback agents.} For the teaching layer: a controlled comparison against the prototype, or against a single-agent baseline, on matched queries. For the feedback layer: comparison of agent-generated interpretations against expert raters who have access to the same underlying data.

\textbf{Bi-directional interactive lectures.} The companion paper~\cite{elhaimeur2026quantum} mentions this briefly. Replacing static video with a streaming voice agent that delivers lecture content while allowing student interruptions introduces a new latency dimension and a new interaction mode. This is system work as much as pedagogical work.

\textbf{Multi-language Code Agent.} The current Code Agent is specialized to Python and Qiskit. Quantum frameworks beyond Qiskit (PennyLane, Q\#, CUDA-Q, diagrammatic languages like Lambeq) would each need their own error catalog and API conventions. Generalizing the Code Agent is primarily a prompt-engineering and data-collection task rather than an architectural one.

\textbf{Larger and more diverse deployment.} Multi-course, multi-instructor deployments would test generalization. Courses where the instructor did not design the system are particularly valuable because self-designed deployments carry bias that the feedback layer cannot correct for.

\section{Conclusion}
\label{sec:conclusion}

This paper described ITAS, a multi-agent LLM-based intelligent tutoring system that was deployed for one semester in a graduate quantum computing course. The architecture has three layers. The teaching layer is a Spoke-and-Wheel of three parallel specialist agents plus a Synthesizer, designed around domain specialization to avoid the task-boundary hallucinations that consolidating video, code, and conceptual reasoning into one agent would risk. The operational layer is four Cloud Run microservices with session state in Cloud SQL and interaction events in BigQuery, designed for classroom-scale concurrency, FERPA compliance, and fast iteration. The feedback layer is a narrow-scope conversational agent that lets the instructor interrogate pseudonymized interaction data without having access to real identities or the ability to write new queries at runtime.

The system carries forward two ideas from the prototype (separation of concerns and multi-agent design) and sets aside two (the knowledge graph and the tag system). The knowledge graph returns in future work when we build the planning layer. The tag system does not: the pacing-inference problem it solved does not exist in the Spoke-and-Wheel design.

The pilot is small and we do not claim learning-outcome results. What we claim is that the architecture described here is one workable answer to the question of what an LLM-based intelligent tutoring system needs to look like end-to-end to run in a real course, and that the evidence from the deployment is consistent with the design decisions we made. The companion papers report the domain-specific findings~\cite{elhaimeur2026quantum} and the performance analysis~\cite{elhaimeur2026latency}; together they describe what we built and how it behaved in the one classroom we ran it in.

\section*{Acknowledgments}
This research was sponsored in part by the Richard T.~Cheng Endowment and supported by Monarch Sphere~\cite{monarchsphere}. Cloud infrastructure was provided through Google Cloud Platform research credits and John D.~Pratt, Seth J.~Hohensee, and Alex L.~Tucker of the ITS Group at Old Dominion University. The QIS curriculum is based on John Watrous's IBM Quantum lecture series. ITAS is developed at the Center for Real-Time Computing (CRTC), Old Dominion University. Gemini was used to improve readability across the article; the authors take full responsibility for all content.

\bibliographystyle{IEEEtran}
\bibliography{ref}

@book{patton2002qualitative,
  author    = {Patton, Michael Quinn},
  title     = {Qualitative Research and Evaluation Methods},
  edition   = {3rd},
  publisher = {Sage Publications},
  address   = {Thousand Oaks, CA},
  year      = {2002},
  note      = {Foundational text on qualitative methodology; argues that validity and
               meaningfulness in qualitative inquiry depend on information-richness
               of cases and analytical capabilities of the researcher, not sample size.}
}

@article{kasneci2023chatgpt,
  author    = {Kasneci, Enkelejda and others},
  title     = {{ChatGPT} for good? {On} opportunities and challenges of large language models for education},
  journal   = {Learning and Individual Differences},
  volume    = {103},
  pages     = {102274},
  year      = {2023},
  doi       = {10.1016/j.lindif.2023.102274}
}

@misc{kim2025scaling,
  author    = {Kim, Yubin and Gu, Ken and Park, Chanwoo and Park, Chunjong and
               Schmidgall, Samuel and Heydari, A. Ali and Yan, Yao and Zhang, Zhihan and
               Zhuang, Yuchen and Malhotra, Mark and Liang, Paul Pu and Park, Hae Won and
               Yang, Yuzhe and Xu, Xuhai and Du, Yilun and Patel, Shwetak and
               Althoff, Tim and McDuff, Daniel and Liu, Xin},
  title     = {Towards a Science of Scaling Agent Systems},
  year      = {2025},
  eprint    = {2512.08296},
  archivePrefix = {arXiv},
  primaryClass = {cs.AI},
  doi       = {10.48550/arXiv.2512.08296},
  note      = {First quantitative scaling principles for LLM-based agent systems.
               Evaluates five canonical topologies across 180 configurations.
               Centralized coordination improves performance by 80.9\% on
               parallelizable tasks and contains error amplification to 4.4x
               vs. 17.2x for independent agents. Capability saturation
               ($\beta=-0.408$, $p<0.001$) shows coordination yields diminishing
               returns once single-agent baseline exceeds $\sim$45\%.}
}

@inproceedings{wu2024autogen,
  author    = {Wu, Qingyun and Bansal, Gagan and Zhang, Jieyu and Wu, Yiran and Li, Beibin and
               Zhu, Erkang and Jiang, Li and Zhang, Xiaoyun and Zhang, Shaokun and Liu, Jiale and
               Awadallah, Ahmed H. and White, Ryen W. and Burger, Doug and Wang, Chi},
  title     = {{AutoGen}: Enabling Next-Gen {LLM} Applications via Multi-Agent Conversation},
  booktitle = {Proceedings of the Third Conference on Language Modeling (COLM 2024)},
  year      = {2024},
  eprint    = {2308.08155},
  archivePrefix = {arXiv}
}

@inproceedings{hong2024metagpt,
  author    = {Hong, Sirui and Zhuge, Mingchen and Chen, Jonathan and Zheng, Xiawu and
               Cheng, Yuheng and Wang, Jinlin and Zhang, Ceyao and Wang, Zili and
               Yau, Steven Ka Shing and Lin, Zijuan and Zhou, Liyang and Ran, Chenyu and
               Xiao, Lingfeng and Wu, Chenglin and Schmidhuber, J\"{u}rgen},
  title     = {{MetaGPT}: Meta Programming for A Multi-Agent Collaborative Framework},
  booktitle = {Proceedings of the Twelfth International Conference on Learning Representations (ICLR 2024)},
  year      = {2024},
  eprint    = {2308.00352},
  archivePrefix = {arXiv}
}

@inproceedings{li2023camel,
  author    = {Li, Guohao and Hammoud, Hasan Abed Al Kader and Itani, Hani and
               Khizbullin, Dmitrii and Ghanem, Bernard},
  title     = {{CAMEL}: Communicative Agents for ``Mind'' Exploration of Large Language Model Society},
  booktitle = {Advances in Neural Information Processing Systems},
  volume    = {36},
  year      = {2023},
  eprint    = {2303.17760},
  archivePrefix = {arXiv}
}

@article{sumers2024coala,
  author    = {Sumers, Theodore R. and Yao, Shunyu and Narasimhan, Karthik and Griffiths, Thomas L.},
  title     = {Cognitive Architectures for Language Agents},
  journal   = {Transactions on Machine Learning Research},
  year      = {2024},
  eprint    = {2309.02427},
  archivePrefix = {arXiv}
}

@inproceedings{guo2024multiagent,
  author    = {Guo, Taicheng and Chen, Xiuying and Wang, Yaqi and Chang, Ruidi and Pei, Shichao and
               Chawla, Nitesh V. and Wiest, Olaf and Zhang, Xiangliang},
  title     = {Large Language Model based Multi-Agents: A Survey of Progress and Challenges},
  booktitle = {Proceedings of the Thirty-Third International Joint Conference on Artificial Intelligence (IJCAI-24)},
  pages     = {8048--8057},
  year      = {2024},
  eprint    = {2402.01680},
  archivePrefix = {arXiv}
}

@article{huang2024hallucination,
  author    = {Huang, Lei and Yu, Weijiang and Ma, Weitao and Zhong, Weihong and Feng, Zhangyin and
               Wang, Haotian and Chen, Qianglong and Peng, Weihua and Feng, Xiaocheng and
               Qin, Bing and Liu, Ting},
  title     = {A Survey on Hallucination in Large Language Models: Principles, Taxonomy, Challenges, and Open Questions},
  journal   = {ACM Transactions on Information Systems},
  year      = {2024},
  doi       = {10.1145/3703155},
  eprint    = {2311.05232},
  archivePrefix = {arXiv}
}

@inproceedings{wang2025genmentor,
  author    = {Wang, Teng and Zhan, Yuzhe and Lian, Jianxun and Hu, Zubin and Yuan, Nicholas Jing and
               Zhang, Qi and Xie, Xing and Xiong, Hui},
  title     = {{LLM}-powered Multi-agent Framework for Goal-oriented Learning in Intelligent Tutoring System},
  booktitle = {Proceedings of The Web Conference 2025 (Industry Track)},
  year      = {2025},
  eprint    = {2501.15749},
  archivePrefix = {arXiv}
}

@misc{chu2025llmagents,
  author    = {Chu, Zhiyuan and Wang, Shichao and Xie, Jingyi and Zhu, Tong and Yan, Yao and
               Ye, Jiaqing and Zhong, An and Hu, Xinzhu and Liang, Jiahan and Yu, Philip S. and Wen, Qingsong},
  title     = {{LLM} Agents for Education: Advances and Applications},
  year      = {2025},
  eprint    = {2503.11733},
  archivePrefix = {arXiv}
}

@inproceedings{zhang2024simclass,
  author    = {Zhang, Zheyuan and Zhang-Li, Daniel and Yu, Jifan and Gong, Linlu and Zhou, Jinchang and
               Liu, Zhiyuan and Hou, Lei and Li, Juanzi},
  title     = {Simulating Classroom Education with {LLM}-Empowered Agents},
  booktitle = {Proceedings of the 2025 Annual Conference of the Nations of the Americas Chapter of the {ACL} (NAACL 2025)},
  year      = {2025},
  eprint    = {2406.19226},
  archivePrefix = {arXiv}
}

@inproceedings{guo2014video,
  author    = {Guo, Philip J. and Kim, Juho and Rubin, Rob},
  title     = {How video production affects student engagement: An empirical study of {MOOC} videos},
  booktitle = {Proceedings of the First {ACM} Conference on Learning @ Scale ({L@S '14})},
  pages     = {41--50},
  year      = {2014},
  doi       = {10.1145/2556325.2566239}
}

@article{blikstein2014programming,
  author    = {Blikstein, Paulo and Worsley, Marcelo and Piech, Chris and Sahami, Mehran and
               Cooper, Steve and Koller, Daphne},
  title     = {Programming pluralism: Using learning analytics to detect patterns in the learning of computer programming},
  journal   = {Journal of the Learning Sciences},
  volume    = {23},
  number    = {4},
  pages     = {561--599},
  year      = {2014},
  doi       = {10.1080/10508406.2014.954750}
}

@article{blikstein2016multimodal,
  author    = {Blikstein, Paulo and Worsley, Marcelo},
  title     = {Multimodal learning analytics and education data mining: Using computational technologies to measure complex learning tasks},
  journal   = {Journal of Learning Analytics},
  volume    = {3},
  number    = {2},
  pages     = {220--238},
  year      = {2016},
  doi       = {10.18608/jla.2016.32.11}
}

@inproceedings{kizilcec2013archetypes,
  author    = {Kizilcec, Ren\'{e} F. and Piech, Chris and Schneider, Emily},
  title     = {Deconstructing disengagement: Analyzing learner subpopulations in massive open online courses},
  booktitle = {Proceedings of the Third International Conference on Learning Analytics and Knowledge ({LAK '13})},
  pages     = {170--179},
  year      = {2013},
  doi       = {10.1145/2460296.2460330}
}

@article{vanlehn2011effectiveness,
  author    = {VanLehn, Kurt},
  title     = {The relative effectiveness of human tutoring, intelligent tutoring systems, and other tutoring systems},
  journal   = {Educational Psychologist},
  volume    = {46},
  number    = {4},
  pages     = {197--221},
  year      = {2011},
  doi       = {10.1080/00461520.2011.611369}
}

@article{kulik2016meta,
  author    = {Kulik, James A. and Fletcher, J. D.},
  title     = {Effectiveness of intelligent tutoring systems: A meta-analytic review},
  journal   = {Review of Educational Research},
  volume    = {86},
  number    = {1},
  pages     = {42--78},
  year      = {2016},
  doi       = {10.3102/0034654315581420}
}

@article{graesser2004autotutor,
  author    = {Graesser, Arthur C. and Lu, Shulan and Jackson, Gordon T. and Mitchell, Heather H. and
               Ventura, Matthew and Olney, Andrew and Louwerse, Max M.},
  title     = {{AutoTutor}: A tutor with dialogue in natural language},
  journal   = {Behavior Research Methods, Instruments, \& Computers},
  volume    = {36},
  number    = {2},
  pages     = {180--193},
  year      = {2004},
  doi       = {10.3758/BF03195563}
}

@inproceedings{elhaimeur2025toward,
  author    = {Elhaimeur, Iizalaarab and Chrisochoides, Nikos},
  title     = {Toward Personalizing Quantum Computing Education: An Evolutionary {LLM}-Powered Approach},
  booktitle = {Proceedings of the IEEE International Conference on Quantum Computing and Engineering (QCE)},
  year      = {2025},
}

@misc{elhaimeur2026quantum,
  author    = {Elhaimeur, Iizalaarab and Chrisochoides, Nikos},
  title     = {From Prototype to Classroom: An Intelligent Tutoring System for Quantum Education},
  year      = {2026},
}

@misc{elhaimeur2026latency,
  author    = {Elhaimeur, Iizalaarab and Chrisochoides, Nikos},
  title     = {Latency and Cost of Multi-Agent Intelligent Tutoring at Scale},
  year      = {2026}
}

@misc{monarchsphere,
  author       = {{Old Dominion University}},
  title        = {{MonarchSphere}: {AI} Incubator powered by {Google Cloud}},
  howpublished = {\url{https://www.odu.edu/forward-focused-transformation/monarchsphere}},
  year         = {2025}
}

@misc{ferpa1974,
  title        = {Family Educational Rights and Privacy Act ({FERPA})},
  author       = {{U.S. Congress}},
  howpublished = {20 {U.S.C.} \S~1232g; 34 {CFR} Part 99},
  year         = {1974}
}

@inproceedings{tack2022aiteacher,
  author    = {Tack, Ana{\"i}s and Piech, Chris},
  title     = {The {AI} Teacher Test: Measuring the Pedagogical Ability of {Blender} and {GPT-3} in Educational Dialogues},
  booktitle = {Proceedings of the 15th International Conference on Educational Data Mining (EDM 2022)},
  pages     = {522--529},
  year      = {2022},
  eprint    = {2205.07540},
  archivePrefix = {arXiv}
}

@inproceedings{tack2023bea,
  author    = {Tack, Ana{\"i}s and Kochmar, Ekaterina and Yuan, Zheng and Bibauw, Serge and Piech, Chris},
  title     = {The {BEA} 2023 Shared Task on Generating {AI} Teacher Responses in Educational Dialogues},
  booktitle = {Proceedings of the 18th Workshop on Innovative Use of NLP for Building Educational Applications (BEA 2023)},
  pages     = {785--795},
  year      = {2023},
  publisher = {Association for Computational Linguistics},
  doi       = {10.18653/v1/2023.bea-1.64},
  eprint    = {2306.06941},
  archivePrefix = {arXiv}
}

@misc{jurenka2024learnlm,
  author    = {Jurenka, Irina and others},
  title     = {Towards Responsible Development of Generative {AI} for Education: An Evaluation-Driven Approach},
  year      = {2024},
  eprint    = {2407.12687},
  archivePrefix = {arXiv},
  note      = {LearnLM tech report; demonstrates that pedagogically fine-tuned Gemini
               (LearnLM-Tutor) is consistently preferred over prompted Gemini baselines
               by educators and learners on multiple pedagogical dimensions.}
}

@misc{vanzo2024gpt4homework,
  author    = {Vanzo, Alessandro and Pal Chowdhury, Sankalan and Sachan, Mrinmaya},
  title     = {{GPT-4} as a Homework Tutor can Improve Student Engagement and Learning Outcomes},
  year      = {2024},
  eprint    = {2409.15981},
  archivePrefix = {arXiv},
  primaryClass = {cs.CY},
  doi       = {10.48550/arXiv.2409.15981},
  note      = {Randomized controlled trial in four high-school classes; reports
               significant gains in grammar learning outcomes and student engagement
               from GPT-4 interactive homework sessions versus traditional homework.}
}

@article{yan2024practical,
  author    = {Yan, Lixiang and Sha, Lele and Zhao, Linxuan and Li, Yuheng and
               Martinez-Maldonado, Roberto and Chen, Guanliang and Li, Xinyu and
               Jin, Yueqiao and Ga\v{s}evi\'{c}, Dragan},
  title     = {Practical and ethical challenges of large language models in education: A systematic scoping review},
  journal   = {British Journal of Educational Technology},
  volume    = {55},
  number    = {1},
  pages     = {90--112},
  year      = {2024},
  doi = {10.1111/bjet.13370},
}

@article{bloom1984twosigma,
  author    = {Bloom, Benjamin S.},
  title     = {The 2 Sigma Problem: The Search for Methods of Group Instruction as Effective as One-to-One Tutoring},
  journal   = {Educational Researcher},
  volume    = {13},
  number    = {6},
  pages     = {4--16},
  year      = {1984},
  doi       = {10.3102/0013189X013006004}
}

@article{anderson1995cognitive,
  author    = {Anderson, John R. and Corbett, Albert T. and Koedinger, Kenneth R. and Pelletier, Ray},
  title     = {Cognitive Tutors: Lessons Learned},
  journal   = {Journal of the Learning Sciences},
  volume    = {4},
  number    = {2},
  pages     = {167--207},
  year      = {1995},
  doi       = {10.1207/s15327809jls0402_2}
}

@inproceedings{yao2023react,
  author    = {Yao, Shunyu and Zhao, Jeffrey and Yu, Dian and Du, Nan and
               Shafran, Izhak and Narasimhan, Karthik and Cao, Yuan},
  title     = {{ReAct}: Synergizing Reasoning and Acting in Language Models},
  booktitle = {Proceedings of the Eleventh International Conference on Learning Representations (ICLR 2023)},
  year      = {2023},
  eprint    = {2210.03629},
  archivePrefix = {arXiv}
}

@misc{googleadk,
  author    = {{Google}},
  title     = {Agent Development Kit ({ADK})},
  howpublished = {\url{https://google.github.io/adk-docs/}},
  year      = {2025},
  note      = {Open-source framework for building, evaluating, and deploying multi-agent
               LLM applications. Used in ITAS for the Spoke-and-Wheel teaching pipeline.}
}

\end{document}